\documentclass[aps,prb,twocolumn,showpacs,floatfix]{revtex4}
\usepackage{graphicx}
\usepackage{color}

\begin{document}

\title{Optical study of superconducting Ga-rich layers in silicon}

\author{T. Fischer}\email{t.fischer@hzdr.de} \author{A. V. Pronin}\email{a.pronin@hzdr.de}
\author{R. Skrotzki} \author{T. Herrmannsd\"orfer} \author{J. Wosnitza}

\affiliation{Dresden High Magnetic Field Laboratory (HLD), Helmholtz-Zentrum Dresden-Rossendorf, 01314 Dresden,
Germany}

\author{J. Fiedler} \author{V. Heera} \author{M. Helm}

\affiliation{Institute of Ion Beam Physics and Materials Research, Helmholtz-Zentrum Dresden-Rossendorf, 01314 Dresden, Germany}

\author{E. Schachinger}

\affiliation{Institute of Theoretical and Computational Physics, Graz University of Technology, 8010 Graz, Austria}

\date{\today}

\begin{abstract}
We performed phase-sensitive terahertz (0.12 -- 1.2 THz)
transmission measurements of Ga-enriched layers in silicon. Below
the superconducting transition, $T_{c}^{\rm middle} = 6.7$ K, we
find clear signatures of the formation of a superconducting
condensate and of the opening of an energy gap in the optical
spectra. The London penetration depth, $\lambda(T)$, and the
condensate density, $n_{s} = \lambda^{2}(0)/\lambda^{2}(T)$, as
functions of temperature demonstrate behavior, typical for
conventional superconductors with $\lambda(0)$ = 1.8 $\mu$m. The
terahertz spectra can be well described within the framework of
Eliashberg theory with strong electron-phonon coupling: the
zero-temperature energy gap is $2\Delta(0)$ = 2.64 meV and
$2\Delta(0) / k_{B}T_{c} = 4.6 \pm 0.1$, consistent with the
amorphous state of Ga. At temperatures just above $T_{c}$, the
optical spectra demonstrate Drude behavior.

\end{abstract}

\pacs{74.25.Gz, 74.70.Ad, 74.78.-w, 74.25.nd}

\maketitle

\section{Introduction}

Recently, some of us observed superconductivity in amorphous Ga-rich
layers manufactured by Ga implantation into Si wafers and subsequent
thermal annealing. \cite{skrotski} Compared to bulk crystalline Ga,
superconductivity occurs at higher temperatures with an onset at 7
-- 10 K. While previously, amorphous Ga films have only been
characterized subsequently to \textit{in situ} low-temperature
synthesis, \cite{buckel, jaeger} the films under investigation
withstand multiple cooling procedures and room-temperature handling.

Here, we report optical evidence for the superconducting-state
formation in gallium-enriched layers in silicon by means of optical
measurements in the terahertz region. These layers consist of
amorphous, gallium-rich precipitates embedded in nanocrystalline
silicon. \cite{fiedler} Optical spectra of superconductors contain
valuable information about the superconducting state: the London
penetration depth, the strength of coupling, the size and the
symmetry of the superconducting gap can be extracted from such
measurements. \cite{tinkham, basov}

In our optical measurements within the terahertz region, we observe
clear signatures of the superconducting condensate developing at $T
< T_{c}$. We were able to trace the temperature dependence of the
spectral weight of the condensate and of the London penetration
depth. We demonstrate that they follow nicely the behavior expected
for fully gaped superconductors. We further show that the
frequency-dependent optical spectra can be well described within the
Eliashberg theory for strong-coupling $s$-wave superconductors.

\section{Experiment}

\begin{figure}[]
\centering
\includegraphics[width=\columnwidth,clip]{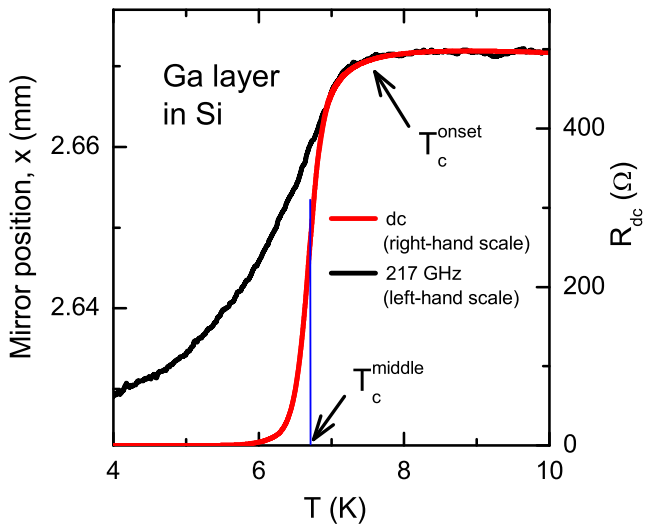}
\caption{(Color online) Temperature dependence of the resistance of
a Ga-enriched layer in Si (right-hand scale) and an example of the
raw measurements of the interferometer mirror position (left-hand
scale). The mirror position is directly related to the optical phase
shift and, hence, to the permittivity and the penetration depth.}
\label{raw}
\end{figure}

Ga$^{+}$ ions had been implanted at an energy of 80 keV with a total
fluence of $4 \times 10^{16}$ cm$^{-2}$ Ga into 0.38 mm thick
(100)-oriented silicon wafers covered with 30 nm of silicon dioxide.
Subsequent rapid thermal annealing at 650 $^{0}$C for 60 seconds has
been applied for realizing gallium precipitation at the Si --
SiO$_{2}$ interface. The thickness of the fabricated Ga-rich layers
was estimated to be 10 nm by means of transmission electron
microscopy and Rutherford backscattering. \cite{fiedler} The samples
probed optically had lateral sizes of 2 by 5 mm. Direct-current
resistivity measurements (Fig. \ref{raw}) show the onset of the
superconducting transition at $T_{c}^{\rm onset} \approx 7.3$ K and
a middle point at $T_{c}^{\rm middle} = 6.7$ K, which we take as
$T_{c}$ for our Eliashberg analysis below. Further information, such
as a detailed structural and critical-field analysis, can be found
in Refs. \onlinecite{skrotski} and \onlinecite{fiedler}.

In the frequency range 3.9 - 41 cm$^{-1}$ (117 - 1230 GHz, 0.48 -
5.1 meV), measurements have been performed by use of a spectrometer,
equipped with backward-wave oscillators (BWOs) as sources of
coherent and frequency-tunable radiation. The measurements have been
done using a number of different BWOs covering the above-mentioned
frequency range almost continuously.

A Mach-Zehnder interferometer arrangement of the spectrometer allows
to measure both the intensity and the phase shift of the wave
transmitted through the sample. A commercial optical $^{4}$He
cryostat with the sample inside is installed in one of the arms of
the interferometer. A mirror in another (reference) arm is connected
to a precision stepper motor (with an accuracy of 0.5 $\mu$m). The
interferometer can be adjusted to a position, where the optical path
difference between the arms is zero. During frequency or temperature
sweeps, a software always keeps the movable mirror in position with
zero optical path difference. Any change in the sample's refraction
index (or dielectric permittivity) results in a change of the
optical path length in the sample arm, and, hence, to a shift of the
movable mirror. This shift is detected and the phase change of the
wave transmitted through the sample is directly calculated from it.
For the amplitude transmission measurements, the reference arm of
the interferometer is blocked with a shutter. The absolute values of
the sample's amplitude and phase-shift transmission, Tr and
$\varphi$, are obtained by repeating the measurements without a
sample and a subsequent correction for the empty-channel
measurement. Fresnel optical formulas \cite{heavens} are used to
extract the optical parameters of the sample (for example, the
complex conductivity, $\sigma = \sigma_{1} + i\sigma_{2}$, or the
complex permittivity $\varepsilon = \varepsilon' + i\varepsilon''$)
from Tr and $\varphi$. A detailed description of the measurement
technique can be found in Ref. \onlinecite{kozlov}. The method has
been previously applied to a large number of different
superconductors. \cite{dressel}

The London penetration depth, $\lambda =
c/(4\pi\sigma_{2}\omega)^{1/2}$ ($\omega$ is the angular frequency,
$c$ is the speed of light), has been determined as a function of
temperature at a number of fixed frequencies (123, 172, 217, and 252
GHz, or 4.1, 5.74, 7.24, and 8.4 cm$^{-1}$) in the same way as
described in Refs. \onlinecite{pronin} and \onlinecite{fischer}. For
these measurements, we limited ourselves to our lowest frequencies,
as at higher frequencies the contribution of normal electrons to
$\sigma_{2}$ becomes significant, thus, a correct determination of
$\lambda$ is difficult.

\begin{figure}[b]
\centering
\includegraphics[width=\columnwidth,clip]{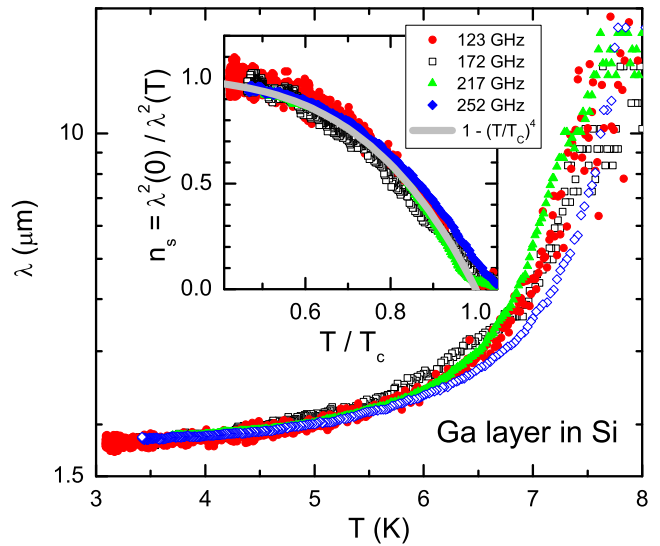}
\caption{(Color online) London penetration depth as a function of
temperature for some frequencies. Inset: Relative density of
superconducting condensate, $n_{s} = \lambda^{2}(0)/\lambda^{2}(T)$,
as a function of temperature for the same frequencies as in the main
panel. Solid line mimics the behavior expected for a fully gapped
superconductor.} \label{lambda}
\end{figure}

The optical parameters of pristine wafers had been measured in
advance using the same setup. At $T \leq 15$ K, we found the real
part of the dielectric permittivity of the wafers to be independent
of temperature and frequency (within our accuracy) and equal to
$\varepsilon' = 11.50 \pm 0.05$. An absorption in the wafers was not
detectable at these temperatures.

\section{Results and discussion}

\subsection{Temperature sweeps. Temperature dependence of penetration
depth and condensate density}

Firstly, let us note that the appearance of superconductivity in our
sample is confirmed by the raw optical data. In Fig. \ref{raw}, we
show an example of such raw data -- the position of the movable
mirror of the Mach-Zehnder interferometer, $x(T)$, as a function of
temperature. Above the superconducting transition, $x(T)$ is flat.
As the sample enters into the superconducting state, $x(T)$ starts
to decrease rapidly. This is due to the condensation of electrons.
The condensed electrons are represented by a delta function in the
real part of the complex conductivity, $\sigma_{1}$. This delta
function leads (\textit{via} the Kramers-Kronig relations) to a
divergence in $\sigma_{2}$ (and in $\varepsilon'$) at $\omega
\rightarrow 0$: $\sigma_{2} = ne^{2}/m\omega$, $\varepsilon' \equiv
1 - 4\pi\sigma_{2}/\omega = 1 - 4\pi ne^{2}/m\omega^{2}$ ($n$ is the
charge-carrier concentration, $e$ is the elementary charge, and $m$
is the effective mass of the carriers). One can easily estimate that
for conventional superconductors at temperatures well below $T_{c}$,
the permittivity at (sub)terahertz frequencies is of the order of
$-10^{3}$ to $-10^{6}$. This negative $\varepsilon'$ leads to a
negative contribution to the phase shift and hence to a negative
contribution to $x(T)$, directly recorded in our measurements.
Obviously, this huge negative $\varepsilon'$ is nothing but
screening of the probing electromagnetic field by supercurrents. As
electrons start to condense already at $T_{c}^{\rm onset}$, $x(T)$
begins to decrease starting at this temperature.

The measured temperature dependences of $\lambda$ for the
above-mentioned frequencies are shown in Fig. \ref{lambda}. All
curves demonstrate a very similar behavior. At low temperatures, the
curves flatten. Extrapolation of $\lambda (T)$ to zero temperature
gives $\lambda (0) = (1.8 \, \pm \, 0.2)$ $\mu$m. As the temperature
approaches $T_{c}$, $\lambda (T)$ nearly diverges for all four
frequencies (because at any non-zero frequency, $\sigma_{2}$ is
different from zero above $T_{c}$, true divergence is not expected).

The inset of Fig. \ref{lambda} shows the relative density of the
superconducting condensate, $n_{s} = \lambda^{2}(0)/\lambda^{2}(T)$,
as a function of temperature. All measurements basically fall onto
one curve, which can be well approximated by a standard
phenomenological expression used to fit the penetration depth data
of conventional (fully  gapped) superconductors: \cite{tinkham}
$n_{s} = 1-(T/T_{c})^{4}$. The highest-frequency curve deviates
slightly from the common trend as $T \rightarrow T_{c}$. We
attribute this deviation to the contribution of normal-state
carriers, as mentioned above (a similar behavior was observed
\textit{e.g.} in Ref. \onlinecite{pronin}). Our measurements are
more noisy at lower frequencies because the wavelength of the
probing radiation at lower frequencies (123 and 172 GHz) becomes
comparable to the lateral sample dimensions.

\begin{figure}[]
\centering
\includegraphics[width=\columnwidth,clip]{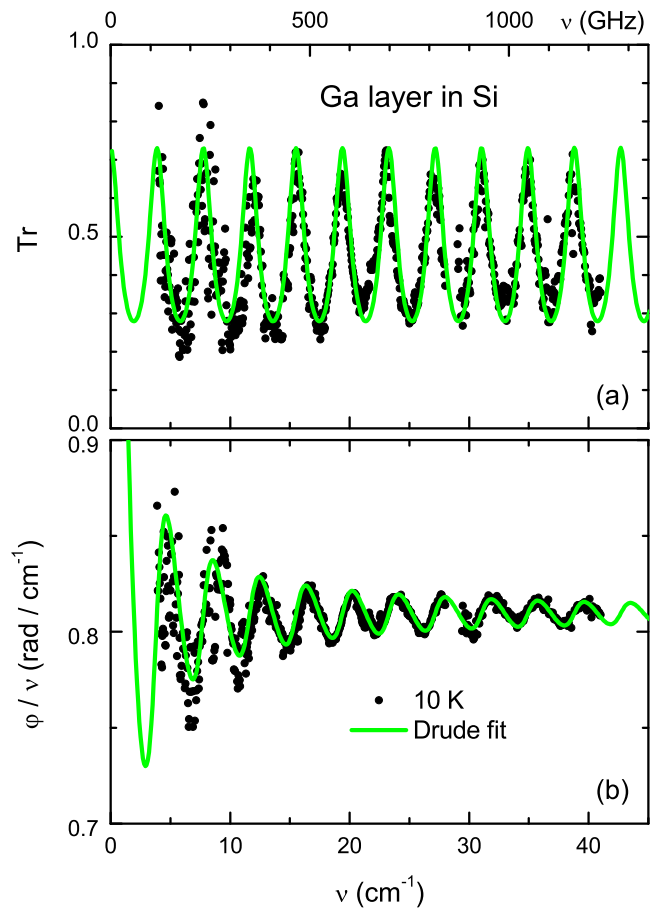}
\vspace{0.0cm} \caption{(Color online) Frequency-dependent
amplitude-transmission (a) and phase-shift (b) spectra of the
Ga-rich layer in Si at $T = 10\,$K. Solid lines show a Drude fit.
The phase-shift spectra are divided by frequency for better
representation.} \label{Tr_Pt_norm}
\end{figure}

\begin{figure}[]
\centering
\includegraphics[width=\columnwidth,clip]{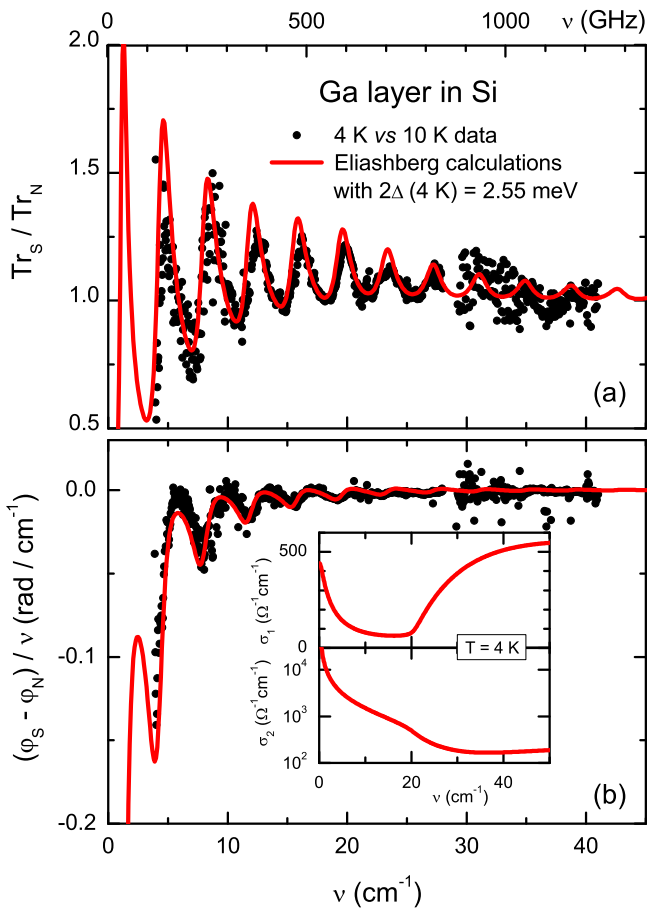}
\vspace{0.0cm} \caption{(Color online) Transmission ratio,
Tr$_{S}$/Tr$_{N}$, (a) and relative phase shift, $\varphi_{S} -
\varphi_{N}$, (b) for the Ga-rich layer in Si as functions of
frequency. Normal-state data are from Fig. \ref{Tr_Pt_norm}.
Superconducting-state data are collected at 4 K. Solid lines show
the Eliashberg calculations for $T = 4\,$K with $2\Delta(4 \rm{K})$
= 2.55 meV = 20.6 cm$^{-1}$. The phase-shift spectra are divided by
frequency. The real, $\sigma_{1}(\nu)$, and imaginary,
$\sigma_{2}(\nu)$, parts of the complex conductivity obtained from
the calculations are shown in the insets. } \label{Tr_Pt_sc}
\end{figure}

\subsection{Frequency sweeps. Drude metal and strong-coupling
$s$-wave superconductor}

\textit{Normal state.} Figure \ref{Tr_Pt_norm} shows (a) the
as-measured transmission, Tr($\nu$), and (b) the phase-shift,
$\varphi(\nu)$, spectra as a function of frequency, $\nu = \omega/2
\pi$ at 10 K. Since the major term of the phase shift is
proportional to the frequency of the probing radiation, the
phase-shift spectra are divided by frequency to eliminate the
constant frequency slope. The pronounced fringes in both Tr($\nu$)
and $\varphi(\nu)$ are due to the multiple interference within the
Si substrate, which acts as a Fabry-Perot interferometer.

The spectra can be well fitted by the Drude model, $\sigma(\omega) =
\sigma_{0}/(1 - i\omega\tau)$, where $\sigma_{0}$ is the
zero-frequency limit of the conductivity and $\tau$ is the
scattering time. We find $\sigma_{0} = (900 \, \pm \, 150)$
$\Omega^{-1} \rm{cm}^{-1}$ in good agreement with our dc data
[$\sigma_{dc} (10 \, \rm{K}) = 1000$ $\Omega^{-1} \rm{cm}^{-1}$].
The scattering rate, $\gamma = 1/(2\pi\tau)$, is above our frequency
range. We can estimate, $\gamma \geq 100$ cm$^{-1}$ ($\tau \leq 5.3
\times 10^{-14}$ s). Lower values of $\gamma$ would result in an
upturn in Tr($\nu$) at the high-frequency end of our range, which we
do not observe here. By use of the Fermi velocity for gallium,
$v_{F} = 6 \times 10^{7}$ cm/s, \cite{gutfeld} we can estimate the
upper limit of the mean free path at 10 K, $\ell \leq 30$ nm.

At our high frequencies, we do not see a downturn of transmission
either. Such a downturn would indicate greater $\sigma_{1}$ values
at higher frequencies and, hence, signal spatial localization of the
carriers. Instead, our Drude-type conductivity proves a
free-electron transport in the Ga layer. From the value of our
highest measurement frequency, we can roughly estimate that no
localization happens on scales of a few dozens of nanometers or
larger. One cannot, of course, exclude localization effects on
smaller spatial scales.

\textit{Superconducting state.} The changes in the spectra related
to superconductivity can best be seen in comparison with the
normal-state spectra.  Therefore, in Fig. \ref{Tr_Pt_sc}, the ratio
between the transmission coefficients at 4 and 10 K,
Tr$_{S}$/Tr$_{N}$, is shown in panel (a); the difference between the
phase shifts, $\varphi_{S} - \varphi_{N}$, measured at the same
temperatures, is in panel (b). The presence of fringes in both
panels indicates a large change of the dielectric constant, which
occurs when the sample enters the superconducting state.

In an attempt to compare experiment with theory we calculated the
complex conductivity within the framework of standard $s$-wave
Eliashberg theory. \cite{carbotte, marsiglio} Details of the
procedure applied in these calculations will be discussed
elsewhere.\cite{fischer_PhD} In agreement with our data on the
temperature dependence of the penetration depth and with the
amorphous state of Ga in our samples, \cite{fiedler} we assumed an
isotropic $s$-wave gap. The Eliashberg function,
$\alpha^{2}(\omega)F(\omega)$, and the Coulomb pseudopotential,
$\mu^{*} = 0.17$ have been taken from Ref. \onlinecite{chen}. The
Eliashberg function was rescaled to give $T^{\rm middle}_c$ keeping
$\mu^\star$ unchanged. This results in a mass-enhancement factor due
to electron-phonon interaction of $\lambda_{el-ph} = 1.97$ down from
2.3 as was reported by Chen {\it et al.}\cite{chen} We obtained the
zero-temperature energy gap: $2\Delta(0) = 2.64$ meV = 21.3
cm$^{-1}$, and $2\Delta(0) / k_{B}T_{c} = 4.6 \pm 0.1$ (the error is
estimated from the width of the superconducting transition). This
value of $2\Delta(0) / k_{B}T_{c}$ is in good agreement with the
results of earlier tunneling and infrared experiments on amorphous
Ga. \cite{cohen, wuehl, minnigerode, harris}

The calculated complex conductivity at $T$ = 4 K is shown in the
insets of Fig. \ref{Tr_Pt_sc} (b). The theoretical complex
conductivity is then used to calculate the transmission ratios and
the phase-shift difference \textit{via} Fresnel formulas [lines in
Fig. \ref{Tr_Pt_sc}]. The good agreement between theory and
experiment is evident.

Finally, from the obtained value for $2\Delta(0)$, we can estimate
the zero-temperature coherence length: $\xi_{0} = \hbar v_{F} / \pi
\Delta(0) \cong 100$ nm. Let us note that although the
Ginzburg-Landau coherence length $\xi^{*}$ derived from the analysis
of the upper-critical-field data, was found to be around 10
nm,\cite{skrotski} there is no contradiction with our result,
because $\xi^{*}$, derived in that way, is strongly affected by the
mean free path.

\section{Conclusions}

Our optical data collected at terahertz frequencies on Ga-enriched
layers in Si: i) support the original observation of a
superconducting transition in resistivity data\cite{skrotski}
\textit{via} the direct detection of the electromagnetic-field
screening due to supercurrents; ii) show that the London penetration
depth [$\lambda (0) = 1.8$ $\mu$m] and the superconducting
condensate density as functions of temperature follow the typical
temperature dependence expected for conventional superconductors;
iii) prove that the frequency-dependent normal-state conductivity is
of Drude type (free electrons) with no signs of localization effects
on length scales equal or larger than several dozens of nanometers;
iv) indicate that the upper limit of the mean free path $\ell$ in
our sample is roughly 30 nm; v) allow to estimate the size of the
superconducting energy gap, $2\Delta(0)$ = 2.64 meV, and the
coherence length, $\xi_{0} = 100$ nm, and to conclude that our
sample is a dirty ($\xi_{0} > \ell$) strong-coupling
($\lambda_{el-ph} = 1.97$) $s$-wave superconductor with $2\Delta(0)
/ k_{B}T_{c} = 4.6 \pm 0.1$.

Part of this work was supported by EuroMagNET II (contract 228043)
and by the DFG (contract HE 2604/7-1).

\end{document}